\documentstyle{article}

\def\be{\begin{equation}}
\def\ee{\end{equation}}
\def\bea{\begin{eqnarray}}
\def\eea{\end{eqnarray}}
\def\({\left(}
\def\){\right)}
\def\[{\left[}
\def\]{\right]}
\def\sign{sign}
\begin{document}

\pagestyle{empty}
\vskip-10pt
\vskip-10pt
\hfill {\tt hep-th/0212167}
\begin{center}
\vskip 3truecm
{\Large\bf
Classical self-dual BPS strings in $d=6$, $(2,0)$ theory from afar
}\\ 
\vskip 2truecm
{\large\bf
Andreas Gustavsson
}\\
\vskip 1truecm
{\it Institute of Theoretical  Physics,
Chalmers University of Technology, \\
S-412 96 G\"{o}teborg, Sweden}\\
\vskip 5truemm
{\tt f93angu@fy.chalmers.se}
\end{center}
\vskip 2truecm
\noindent{\bf Abstract:} 
We show how one can get solitonic strings in a six-dimensional (2,0) supersymmetric theory by incorporating a nonlinear interaction term. We derive a zero force condition between parallel strings, and compute a metric on a moduli space which is $R^4$ when the strings are far apart. When compactifying the strings on a two-torus we show that, in the limit of vanishing two-torus, one regains the moduli space of two widely separated dyons of equal magnetic charges in four dimensions.

\vfill \vskip4pt

\eject
\newpage
\pagestyle{plain}

\section{Introduction}
We will in this letter study the zero modes of tensile BPS strings in an interacting (2,0) theory in five space plus one time dimension. By zero modes we will mean modes that do not change the energy. We will thus not consider excited strings here. We show that it is possible to have such strings in a theory as solitons. The model we consider thus contains no fundamental strings. The only fields we consider are the massless ones which are in the (2,0) tensor multiplet, and which includes a self-dual tensor field $H^+$ plus five scalars $\phi^a$, $a=1,...,5$. The strings being tensile means that we study the theory away from the origin of some M-theory moduli space, corresponding to separated M5-branes. We thus break the conformal invariance by giving non-zero expectation value to one of the five scalar fields. We then can expand the quantum fields about such classical solitons in the spirit of \cite{Jackiw-Rebbi}. This should presumably work if we consider energies much lower than $D^{\frac{1}{2}}$ as the work done in \cite{Henningson-Flink} may indicate. Here $D$ is the expectation value of the fifth scalar field at infinity. We can not make an expansion in a small coupling constant since we do not have any adjustable coupling constant in these six-dimensional theories.

The structure of this paper is as follows. In section 2 we present the model. In section 3 we show how to obtain charged solitonic BPS strings. In sections 3 we show that these strings can have fermionic zero modes, thus obtaining spinning strings. In section 4 we examine the bosonic zero modes of BPS strings and compute the metric on the moduli space. In section 5 we make contact with four dimensional results.

\section{The model}
We will throughout this letter assume that we have a flat six-dimensional Minkowski metric $\eta_{MN} = diag(-1,1,1,1,1,1)$. We will take as our six-dimensional action the following
\be
\frac{1}{2\pi}\int_W \left(-\frac{1}{2}H\wedge *H - d\phi^a\wedge *d\phi_a + H\wedge A(\phi^a)\) + fermions,
\ee
where we will express the three-form potential $A$ (coming from eleven dimensional supergravity, having field strength $F = dA$) in terms of the five scalars living on the brane. That the term $\int H\wedge A$ should be there in the eleven dimensional context has been shown in \cite{Aharony}. In \cite{Strominger} this term was used to show that an M2 brane can end on an M5 brane. We will assume that we have two parallel M5 branes, and we will factorize out the `center of mass' tensor multiplet, so that we get the (spontaneously broken) $A_1$ theory described by one massless tensor multiplet (plus massive degrees of freedom, which we do not know how to describe properly).

We will now examine the effects that the second five brane $W'$ have on $W$. Small fluctuations of $W'$ can presumably be neglected when the two M5 branes are widely separated. But if there is a M2 brane stretching between the two M5 branes, then we can not trust such a model when being close to the boundary of such a M2 brane, which will be our selfdual string in the M5 brane. But far away from such strings we should be able to solve $A$ to a good approximation from 
\be
dF = 2\pi \delta_{W'}
\ee
with $\delta_{W'}$ being the Poincare dual of a flat five-brane world volume $W'$. It has the solution $F = 2\pi \frac{1}{V_4}\Omega_4$ where $\Omega_4$ and $V_4$ are the volume form and volume respectively of the unit four-sphere that surrounds $W'$. If we let 
\bea
\phi^5 & = & r \cos \theta\cr
\phi^4 & = & r \sin \theta \cos \phi \cr
\phi^3 & = & r \sin \theta \sin \phi \cos \psi\cr
\phi^2 & = & r \sin \theta \sin \phi \sin \psi \cos \chi\cr
\phi^1 & = & r \sin \theta \sin \phi \sin \psi \sin \chi
\eea
denote the five transverse coordinates to $W'$, then
\bea
\Omega_4 & = & \frac{1}{4!}\epsilon_{abcde}\frac{1}{r^5}\phi^a d\phi^b \wedge d\phi^c \wedge d\phi^d \wedge d\phi^e \cr
& = & \sin^3 \theta \sin^2 \varphi \sin \psi d\theta\wedge d\varphi\wedge d \psi \wedge d\chi\cr
& = & \sin^3 \theta d\theta \wedge \Omega_3.
\eea
Here $\Omega_3$ is the volume form on the equator $\phi^5 = 0$. Locally $F = dA$. We may for instance choose $A = \frac{2\pi}{V_4} a(\theta)\Omega_3$ for some function $a(\theta)$ such that $da = \sin^3 \theta d\theta$. Such solutions can only be obtained locally by excluding either the south pole ($\theta = \pi$) or the north pole ($\theta = 0$). They are given respectively by
\be
A_{\pm} = \frac{2\pi}{V_4} \left(\pm \frac{2}{3} - \cos \theta + \frac{1}{3} \cos^3 \theta\right) \Omega_3.
\ee
In Cartesian coordinates (with $A, B, ... = 1,2,3,4$) we get
\bea
A_{\pm} & = & \frac{\pi(\phi^5 \pm 2r)}{3 V_4 r^3(\phi^5 \pm r)^2}\varepsilon_{ABCD}\phi^A d\phi^B\wedge d\phi^C\wedge d\phi^D,
\eea
so we see that $A_{-}$ ($A_{+}$) is singular along the positive (negative) $\phi^5$-axis.

In order for the anti-selfdual piece of $H$ to decouple we should define the field strength as \cite{Witten96}
\be
H = dB + A(\phi)
\ee
where $B$ is the two-form gauge potential. This definition implies the Bianchi identity 
\be
dH = F(\phi).
\ee
The equations of motion one derives from this action are
\bea
d*H & = & \frac{2\pi}{4!V_4}\epsilon_{abcde}\frac{1}{r^5}\phi^a d\phi^b \wedge d\phi^c \wedge d\phi^d \wedge d\phi^e\cr
d*d\phi_a & = & \frac{\pi}{3!V_4} \epsilon_{abcde}\frac{1}{r^5}(H + *H) \wedge \phi^b d\phi^c \wedge \phi^d \wedge d\phi^e.
\eea
These equations have also been considered in \cite{Motl}. We see that only $H^+ = \frac{1}{2}(H + *H)$ couples to the current $F(\phi)$. The equations for the five scalars can be derived as follows. The variation of the Lagrangian is
\be
\frac{1}{2\pi}\left[2\delta \phi^a d*d\phi^a - \delta A(\phi) \wedge (*H + H)\right].
\ee
We choose the gauge corresponding to $A_-$ and define $\rho = \sqrt{\phi^A \phi_A}$ ($A = 1,2,3,4$) and compute the variation with respect to $\phi^5$,
\be
\delta A(\phi)  =  \frac{2\pi}{V_4}\delta(a(\theta)\Omega_3)
 =  \frac{2\pi}{V_4}\sin^3 \theta \delta \theta \Omega_3
 =  - \frac{2\pi}{V_4}\sin^2 \theta \delta (\cos \theta) \Omega_3
 =  - \frac{2\pi}{V_4}\delta \phi^5 \frac{1}{r^5} \rho^4 \Omega_3.
\ee
Hence the corresponding equation of motion is
\be
2 d * d \phi^5 = \frac{2\pi}{V_4} (H + *H)\wedge \frac{1}{r^5} \rho^4 \Omega_3.
\ee
where $\rho^4 \Omega_3 = \frac{1}{3!}\epsilon_{5bcde}\phi^b d\phi^c \wedge d\phi^d \wedge d\phi^e$. The equation of motion for any other $\phi^a$ is most easily derived by choosing the gauge in which $A$ is singular along the positive $\phi^a$-axis. The equations of motion are of course gauge invariant, so this is a legitimate way to proceed.

We notice that the term $H\wedge A(\phi)$ in the Lagrangian depends linearly on $\dot{B}$. This implies that this term vanishes under a Legendre transformation. If we choose the gauge $A_{0IJ} = B_{0I} = 0$ ($I,J = 1,2,...,5$) then the conjugate momenta of $B_{IJ}$ become
\be
2\pi E^{IJ} = -\frac{1}{2}H^{0IJ}
\ee
Letting $H^{\pm} = \frac{1}{2}(1 \pm *)H$, the bosonic Hamiltonian in the rest frame becomes 
\bea
2\pi M & = & \int \left({H^+}\wedge *_5 {H^+} + {H^-}\wedge *_5 {H^-} + d\phi^a \wedge *_5 d\phi^a \right).
\eea
Rewriting the mass as \cite{Townsend}
\be
2\pi M = \int d^5 x \(\left|H^+ \pm *_5\left(d\phi^5\wedge dx_5\right)\right|^2 + (\partial_5 \phi^5)^2 + \left|H^-\right|^2 + |d\phi^A|^2\) \mp 2\int H^+ \wedge d\phi^5 \wedge dx_5,
\ee
we obtain a lower bound on the mass given by
\be
2\pi M \geq 2 \left|\int H^+ \wedge d\phi^5 \wedge dx_5 \right|.
\ee
This is not the most general lower bound that we can have. The situation that we have described is the bound that corresponds to a single selfdual string in the $x^5$-direction. In a supersymmetric theory we know that this bound should be given through a central charge in the supersymmetry algebra. Although the supersymmetry generators may have to be corrected by complicated nonlinear terms in order to be able to close the supersymmetry algebra on the new shell that we get when adding nonlinear interaction terms to the action, we were able to deduce that the central charge one gets still should be given by the simple expression that one derives in the free theory, namely $Z_{\mu}^a = 2\int H^+ \wedge d\phi^a \wedge dx_{\mu}$. In the bound above this central charge has rank one (viewed as a matrix in the indices $\mu$ and a). In general it could have higher rank, corresponding to several non-parallel strings. Now if we choose the central charge as $2QD\int dx^5$ with $D>0$ we get the following Bogomolnyi equations which have also been analysed in \cite{Lambert},
\bea
H^+ & = & \sign(Q) *_4 d\phi^5\cr
d\phi^A & = & 0\cr
\partial_5 \phi^5 & = & 0\cr
H^- & = & 0.
\eea
Of course we can always rotate the coordinate system so that the rank one central charge will be aligned in the $x^5$-direction.

\section{A classical BPS string solution}
We will look for a BPS solution where the scalar field $\phi^5$ is spherically symmetric, i.e. depend only on $R = \sqrt{x^A x_A}$, ($A = 1,2,3,4$). All fields are assumed to be translationally invariant in the $x^0,x^5$ directions (which we will associate to the directions of a string world-sheet). We will assume the asymptotic behaviour
\bea
\phi^5(\infty) & = & D > 0\cr
\phi^A(\infty) & = & 0\cr
\phi^A(0) & = & 0.
\eea
Of course it now follows from the Bogomolnyi equations that $\phi^A = 0$ everywhere. But in that case we can not find any non-trivial solution. We therefore first look for a charged solution, which is not necessarily BPS. The fact that the solution is charged and thereby topologically non-trivial prevents it from becoming trivial. So then we can take the limit that $\phi^A$ tends to zero to obtain a solution that is BPS.
 
We may integrate up the equation of motion $dH = F$ if $A$ is globally defined on $W$ by using Stokes theorem. If the map $x^A \mapsto \frac{\phi^A}{\rho}$, $\rho = \sqrt{\phi^A\phi^A}$, has winding number $N$ then we get
\be
\int_{S^3} H = \frac{2\pi}{V_4}\(-\frac{2}{3} - \cos \theta + \frac{1}{3} \cos^3 \theta\) N V_3.
\ee
We have thus assume that the point $(\rho,\phi^5) = (0,D)$ never is reached for any finite $R$ and that $\phi^5<0$ at $R=0$ where $\rho=0$. This means that we may use the potential $A_-$ which is singular on the positive $\phi^5$-axis. This potential will then be non-singular all over the M five-brane world-volume. Furthermore, if we compute the charge $Q$ by integrating $H$ over a three-sphere with infinite radius $R$ we get $Q = \int_{S^3_{\infty}} H = -2\pi N$. We can now consider a limiting process which will not alter this topological twist to obtain a BPS configuration. We may for instance take the square integrable function $\rho = \epsilon R e^{-R}$ and let $\epsilon \rightarrow 0$. Using the Bogomolnyi equation which relates $H$ and $d\phi^5$ we get,
\be
R^3 {\phi^5}'(R) = |Q| \frac{1 + \sign(\phi^5)}{2V_3} + Ordo(\epsilon^4).
\ee
In the limit $\epsilon \rightarrow 0$ we then get the BPS solution
\bea
\phi^5 & = & \(D - \frac{|Q|}{2V_3 R^2}\)\Theta\(R-\sqrt{\frac{|Q|}{2V_3D}}\)\cr
\phi^A & = & 0.
\eea
Here $\Theta$ is the Heaviside step function. This is thus the exact solution in our model. But our model is not realistic when $\phi^5$ is close to zero. The fact that charge quantization comes about from the topological twist of the scalars around the string should however be true no matter how many further interaction terms we add to our action to make it more realistic. One argument for why one should think so, apart from the obvious one that we got the correct Dirac charge quantization in this way, is that the charge should be a topological invariant (since it can not jump under a continuous deformation) and we may deform our BPS solution continuously in such a way that the two M5 branes never come close to each other. We can also deform $W'$ away from its true stable configuration into the flat brane that we have based our model on without changing the topology. In that situation we may rely on our model. That the charge corresponds to a winding number of the scalars has been anticipated earlier \cite{Motl}, \cite{Intriligator}, but perhaps not been shown this explicitly.

We may insert this solution into $dH$ to compute $F$. We then find that
\be
dH = Q \delta\(R-\sqrt{\frac{|Q|}{2V_3D}}\)dR \wedge \frac{\Omega_3}{V_3}\label{BPS}
\ee
If we have two strings far away from each other then we should be able to superimpose two source terms like that above, with one such source term centered at the each string. So although the equations we started with were non-linear, when restricting to BPS solutions the equations become linear and the superposition principle holds, at least when the strings are widely separated.

\section{The fermionic zero modes}
The equation of motion for the fermions can be obtained by making a supersymmetry transformation of the bosonic equations. Imposing the supersymmetry variations (which should be correct up to linear order in the fields)
\bea
\delta B_{MN} & = & \epsilon\gamma_{MN}\psi\cr
\delta \phi^a & = & \epsilon \sigma^a \psi \label{susy1}
\eea
(where $\gamma^M$ and $\sigma^a$ are gamma-matrices, $\epsilon$ a constant anticommuting symplectic Majorana spinor parameter and $\psi$ the symplectic Majorana spinor field which is in the $(2,0)$ tensor multiplet) we get from the supersymmetry variation of the bosonic equations of motion that
\be
\gamma^{M} \partial_{M} \psi = \frac{\pi}{75 V_4} \epsilon_{abcde}\frac{1}{r^5}\phi^a \partial_{M}\phi^b \partial_{N}\phi^c \partial_{P}\phi^d \gamma^{MNP} \sigma^e \psi.
\ee 
The right-hand side will be a source term similar to (\ref{BPS}) for BPS strings. If we consider a soliton in which all fermionic fields are put equal to zero, then of course the supersymmetry variations of the bosonic fields vanish. The supersymmetry variation of the fermions is (to linear order)
\be
\delta \psi = -\(\frac{1}{24}\gamma^{MNP}H^+_{MNP} + \frac{1}{2} \sigma_a \gamma^{M}\partial_{M}\phi^a\) \epsilon\label{susy2}.
\ee
When inserting our classical bosonic soliton solution into this variation (thus following the same semi-classical arguments as in \cite{Osborn}) we obtain four (complex) fermionic zero modes (corresponding to the eight real supersymmetries which are broken by our string soliton),
\be
\psi_0 = \frac{1}{\sqrt{|Q|DV_3\int dx^5}}\gamma^{05I}H_{05I}u
\ee
($I = 1,2,3,4$) where $u$ is a constant spinor that survives the projection $P = \frac{1}{2}(1 + \sigma_5 \gamma^0\gamma^5)$. This is a projector onto the broken supersymmetries. There are four complex solutions $u$ that survives this projection. These are static solutions of the equation of motion and hence zero modes, but in general there could exist more fermionic zero modes. It seems plausible that for winding numbers $\pm 1$ of the scalars around the string, i.e. for the charges $Q = \pm 2\pi$, these are all the fermionic zero modes. Expanding $\psi = \psi_0 $ + (non-zero modes) in an orthonormal basis, the equal-time canonical quantization relations $\{\psi(x),\psi(y)^{\dag}\} = \delta^5(x-y)/4$ (together with the completeness and orthonormality relations of the basis) implies that $\{u,u^{\dag}\} = 1/4$. We may write $u = a_i u_i$ where $a_i$ are operators obeying $\{a_i,a_j^{\dag}\} = \delta_{ij}/4$ and $u_i$ ($i=1,...,4$) are c-number valued constant complex orthonormal spinors.

We may insert our classical static soliton solution into the supercharges (they are given in \cite{Gustavsson} in terms of the fields), and find that the supercharges are equal to $4\sqrt{|Q|D\int dx^5}Pu$, so the supersymmetry algebra will be obeyed. Since the Hamiltonian commutes with the supercharges we can construct new string states which have the same energy by acting on the soliton state with vanishing $SO(4)$-spin \footnote{The string breaks the Lorentz group $SO(1,5)$ down to $SO(4)$} by any one of the following 16 creation operators $1,a^{\dag}_i,a^{\dag}_i a^{\dag}_j, a^{\dag}_i a^{\dag}_j a^{\dag}_k, a^{\dag}_1 a^{\dag}_2 a^{\dag}_3 a^{\dag}_4$ to obtain the short BPS multiplet of spinning string states that was constructed in \cite{Henningson-Gustavsson}.

\section{The moduli space of two bosonic strings}
The minimal tension (we assume that we have translational invariance in the $x^0$ and $x^5$ directions) in a system with fields $H_T$ and $\phi_T$ is given by
\be
2\pi T = 2\left|\int_{R^4} H_T \wedge d\phi_T\right|
\ee
and this is so only if $H_T=\pm *d\phi_T$ for some sign choice. Now if these fields are produced by a system of two strings, each of which give rise to the fields $H$ and $\tilde{H}$ where $\tilde{H}(x) = cH(x - R)$, c being some constant, is the field produced by the second string separated from the first string by a distance $\vec{R}$, then since the superposition principle holds outside the strings where the source terms vanish, we get the total field as $H_T = H + \tilde{H}$. Charge quantization requires $c$ to be a rational number. Similarly $d\tilde{\phi} = c' d\phi$ for some constant $c'$. The two strings are BPS separately only if
\bea
H & = & \sign(Q)*d\phi\cr
\tilde{H} & = & \sign(\tilde{Q})*d\tilde{\phi}.
\eea
Hence $c'=sign(Q)sign(\tilde{Q})c$. The whole system will then in general not be BPS since now we get
\be
H_T = \sign(Q)*d\phi + \sign(\tilde{Q})*d\tilde{\phi}
\ee
and this is equal to $\sign(\tilde{\tilde{Q}})(*d\phi + *d\tilde{\phi})$ for some $\tilde{\tilde{Q}}$ (which will be equal to $\tilde{Q} + Q$), as is necessary for the configuration to be BPS, if and only if $\sign(Q) = \sign(\tilde{Q})$. 

The tension of one of the strings we define as follows. We first notice that
\be
\int H_T \wedge d\phi_T = \int H\wedge d\phi_T + \int \tilde{H}\wedge d\phi_T 
\ee
and that we may compute these integrals by performing the integrations as follows,
\be
\int H \wedge d\phi_T = \int d\phi_T \int_{\phi_T = const} H
\ee
The level curve $\phi_T = const$ will for sufficiently small $\phi_T$ consist of disconnected components, each of which encloses just one string. But only that closed contour that surrounds the string that is producing the field $H$ will give a contribution to the integral $\int_{\phi_T = const} H$ and this will have the value $Q$. This value is independent of $\phi_T$. So all in all we get the result
\be
\int H_T \wedge d\phi_T = \sum_i Q_i D
\ee
where the sum runs over all strings, having charges $Q_i$ respectively, and $D = \lim_{R\rightarrow \infty} \phi_T$. It is now natural to interpret $2|Q_i| D$ as the tension of string number $i$ in this configuration. (The factor $2$ is just a convention.)

Furthermore, the contributions $D_i = \lim \phi$ from each string separately must add up to $D$. From $c' = c$ it follows that $D_i = \frac{|Q_i|}{|\sum_j Q_j|}D$.

As a consequence of the zero force property of BPS configurations we have a moduli space for two parallel BPS strings. Since we found four fermionic zero modes we expect of a supersymmetric theory to find exactly four bosonic zero modes as well (and not more, if we ignore center of mass motion). We therefore expect the asymptotic moduli space to be $R^4$ where a point in $R^4$ corresponds to the relative separation of the two strings. It might not be so interesting to scatter infinitely long and therefore infinitely heavy strings against each other. But if we compactify the direction along which they are stretching on a circle then their masses become finite, and it could be of interest to compute the metric on this moduli space. In order to do that we first notice that the bosonic fields produced by a string, outside that region where the fields are zero, happens to be, in our model, exactly reproducible by an action like\footnote{If this action also can be used for fluctuating strings is not clear, but we will not consider fluctuating strings in this letter. As the strings appears to look like singular objects in this action, one might wonder if these solitonic strings really could be fundamental objects. But as the strings are self-dual nothing essential changes under electric-magnetic duality. There is no way to escape the magnetic coupling. In Yang-Mills theory that would have implied that such objects would have to be solitonic. But of course this is not a Yang-Mills theory, so we should not exclude any possibilities.}
\bea
&&-\int \frac{1}{2}H\wedge *H - |Q|\int_{\Sigma} B\cr
&&-\int d\phi \wedge *d\phi - 2|Q|\int_{\Sigma}d^2\sigma\sqrt{-g}\phi
\eea
where $\Sigma$ is the string world-sheet oriented such that $\int_{R^4} \delta_{\Sigma} = sign(Q)$, and $H = dB + |Q|\delta_D$ with $D$ being a Dirac brane extending from $\Sigma$ out to infinity. The induced metric on $\Sigma$ is denoted $g_{ab}$ and $g = det(g_{ab})$. The peculiar normalization of the kinetic term for the scalars follows from our conventions for the supersymmetry variations \cite{Gustavsson}. In this action there is implicitly a coupling term to the magnetic charge as well, coming from\footnote{Our definition of Poincare dual is $\int \omega\wedge \delta_D = \int_D \omega$. Then we must define $\partial D = -\Sigma$ if we want $d\delta_D = \delta_{\Sigma}$}
\be
-\int \frac{1}{2}H\wedge *H = -\int \frac{1}{2}dB\wedge *dB - |Q|\int \delta_D \wedge *H = -\int \frac{1}{2}dB\wedge *dB + |Q|\int_D *H.
\ee
If we define the dual potential $\tilde{B}$ as $d\tilde{B} = *dB$, then the last term can be written as 
\be
-|Q|\int_{\Sigma}\tilde{B}.
\ee
Due to self-duality we can on-shell simply choose $B=\tilde{B}$ when we are away from the Dirac-brane $D$, and we may simply collect the coupling terms to a term that is twice the electric coupling term. It is now important to note that this action, in which we have used the self-duality of $H$, then can be used only to derive the equations of motion for the strings which move in a given on-shell $B$-field. It can then not be used to derive the equations for the $B$-field.

As we chose the orientations according to the sign of the charges, this is exactly the same as saying that the whole system is BPS if and only if the two strings are parallel. But $Q$ and $\tilde{Q}$ need not be equal. It suffices that they have same sign. But if $Q = 2\pi N$ for some $N \neq \pm 1$, then we think that we can view this as $N$ copies of a string with $N = \pm 1$ which are on top of each other, and which then could be separated from each other without any energy cost. So we will in the sequel assume that $N = \pm 1$.

We now set out to compute the metric on the asymptotic part of the moduli space which is $R^4$ by following the same method as in \cite{Manton}. The fields produced by a string can be expressed in terms of a Greens function $J(x)$ as follows
\bea
B_{MN}(x) & = & |Q|\int_{\Sigma} dX_{M}\wedge dX_{N} J(x - X) + |Q|\epsilon_{MNPQRS}\int_{D}dY^{P}\wedge dY^{Q}\wedge dY^{R} \frac{\partial}{\partial x^{S}} J(x - Y)\cr
\phi(x) & = & |Q|\int_{\Sigma} d^2\sigma \sqrt{-g} J(x - X(\sigma)).
\eea
where $J(x)$ satisfies $\partial^M\partial_M J(x) = \delta^6(x)$. The retarded solution to this equation is
\be
J(x^0,r) = -\frac{1}{3V_3}\[\frac{1}{r^3}\delta(x^0 - r) + \frac{1}{r^2} \delta'(x^0 - r)\]
\ee
where $r = \sqrt{R^2 + (x^5)^2}$ and $R = \sqrt{(x^1)^2 + (x^2)^2 + (x^3)^2 + (x^4)^2}$. The potential above will be singular on the Dirac-brane $D$.

If string number one moves with a velocity $v_1$ which is much less than one (the speed of light), and we parameterize $\Sigma_1$ as
\bea
X^0 & = & \tau\cr
X^I & = & {v_1}^I \tau\cr
X^5 & = & \sigma
\eea
and the associated Dirac brane $D$ as
\bea
Y^0 & = & \tau\cr
Y^{1,2,3} & = & {v_1}^{1,2,3} \tau\cr
Y^{4} & = & \tilde{\sigma} + {v_1}^4 \tau\cr
Y^5 & = & \sigma,
\eea
then the scalar field produced by $\Sigma_1$ becomes
\be
\phi = D_1 - \sqrt{1-\vec{v_1}^2}\frac{|Q_1|}{2V_3 S^2} + Ordo((v_1)^3)
\ee
where $S=R(1+Ordo(v_1))$. Similarly, the retarded potential becomes
\bea
B_{05} & = & -\frac{Q_1}{2V_3 S^2} + Ordo((v_1)^3)\cr
B_{IJ} & = & \omega_{IJ} + Ordo(v_1)\cr
B_{0J} & = & v^I \omega_{IJ} + Ordo((v_1)^2)\cr
B_{I5} & = & {v_1}_I B_{05}.
\eea
Here $\omega_{MN}$ is the potential generated by a static magnetically charged string with Dirac-brane singularity along $D$. We assume that we can neglect the accelaration. It is then exactly the {\sl{same}} quantity $S$ that enters in all these expressions. This is a crucial observation.

The expressions above which involve $\omega$ turn out to be irrelevant for us since these components are not in the direction of $\Sigma$. (This is the essential difference as compared to the four dimensional case, which there gave an interpretation of electric charge as a fourth moduli \cite{Manton}. Here we are lucky that we do not get any contribution from the magnetic components as we do not want any more moduli than the four that we already have found.)

Notice that we can make periodic $x^5$ and still get the same retarded solutions as above because that solution is constant in the $x^5$ direction and hence satisfy both the correct equations of motion as well as any periodic boundary conditions that we might wish to impose in the $x^5$-direction.

Inserting these fields into the action for a second string parallel to the first (which implies that their charges have the same sign) and at a distance $\vec{R}$ from it, we get (now we can replace $S$ by $R$ as we see that the $Ordo(v_1)$-corrections cancel, so that the effect of replacing $S$ by $R$ will only give corrections of $Ordo({v_1}^3)$) 
\bea
&&|Q_2|\(-2\int d^2\sigma \sqrt{-g} \phi - 2\int B\) \cr
& = & 2|Q_2|\int d^2 \sigma \[-D_1\sqrt{1-{\vec{v_2}}^2} + \frac{|Q_1|}{R^2} \(-\sqrt{1-{\vec{v}_1}^2}\sqrt{1-{\vec{v}_2}^2} + 1 + \vec{v}_1 \cdot \vec{v}_2 \) \].
\eea
Upon adding the kinetic term of $\Sigma_2$, which has $D_2 = \frac{|Q_2|}{|Q_1 + Q_2|} D$, the center of mass motion decouples and the action governing the relative motion is given by
\be
\int d^2\sigma \(\frac{Q_1 Q_2}{2|Q_1 + Q_2|}D - \frac{Q_1 Q_2}{\vec{R}\cdot\vec{R}}\) \dot{\vec{R}}\cdot\dot{\vec{R}} 
\ee
The strings will move in geodesics on this $R^4$ with metric
\be
(ds)^2 = \(\frac{Q_1 Q_2}{2|Q_1 + Q_2|}D - \frac{Q_1 Q_2}{\vec{R}^2}\) dR^i dR^i
\ee
where $\vec{R}$ is the relative separation between the strings. This is a valid approximation only when $|\vec{R}|>> D^{-\frac{1}{2}}$.

\subsection{Reduction to four dimensions}
The moduli space of these strings should be connected to the moduli space in four dimensions . At least we should be able to reduce to the moduli space for two monopoles that are far away from each other by following the route in \cite{Manton}. We can get the four-dimensional action that was considered in \cite{Manton} by compactifying $x^4$ and $x^5$ on a small $T^2$. We thus make the identifications $x^4 \sim x^4 + q R$ and $x^5 \sim x^5 + g R$. We consider two parallel strings on $T^2$ which are aligned along the `diagonal' (that is, they are straight, and wind around both the cycles arbitrarily many turns). We parameterize the string world-sheets $\Sigma_i$ $(i = 1,2)$ as follows
\bea
X^0 & = & \tau\cr
X^i & = & v^i \tau\cr
X^4 & = & n R q\sigma + v^4 \tau\cr
X^5 & = & m R g\sigma,
\eea
$0\leq \sigma \leq 1$ and $n, m$ being integers. (Thus they are winding numbers of the string around the a- and b-cycles). We will call such a string an (n,m)-string. The four-dimensional electric and magnetic potentials should now be defined as
\bea
A_{\mu} & = & R B_{\mu 4}\cr
\tilde{A}_{\mu} & = & R B_{\mu 5}
\eea
($R$ is assumed to have dimension length, and we choose $q$ and $g$ such that $qg = 2\pi \hbar$). We will only care about the coupling terms to the string (or particle upon dimensional reduction) and will not worry about getting a the Lagrangian whose variation with respect to $B$ yields the correct equation of motion for the gauge field. Instead we will assume that the gauge field already satisfies the equation of motion. That is, we assume that $H$ is self-dual. That will in turn relate $A$ and $\tilde{A}$ as each others dual potentials.

If we let $\Sigma = \gamma \times S^1$ where $\gamma$ thus describes the trajectory of $X^{\mu}(\tau)$ for $\mu = 0,1,2,3$ and $S^1$ describes the trajectory of the coordinates $X^{4,5}(\tau,\sigma)$ around the $T^2$, then the six-dimensional coupling of a string to an on-shell gauge field reduces to 
\be
\int_{\Sigma} B = \int_{\gamma} \left\{n q A + m g \tilde{A} + v^4 nm qg R^2 B_{45} \right\}.
\ee
The last term is nothing that we recognize, but it will turn out to be irrelevant when we take the limit $R \rightarrow 0$.

If we use the notation $\dot{X}^{\mu} = \frac{\partial {X}^{\mu}}{\partial \tau}$, ${X'}^{\mu} = \frac{\partial{X}^{\mu}}{\partial \sigma}$, $(\dot{X})^2 = \dot{X}_{\mu}\dot{X}^{\mu}$, and let $a, b, ... = 0,1$ correspond to $\tau$ and $\sigma$ respectively, then the induced metric on the world-sheet becomes
\be
g_{ab} = \left(
\begin{array}{cc}
(\dot{X})^2 + (\dot{X}^4)^2 & \dot{X}^4{X'}^4 \\
\dot{X}^4{X'}^4 & ({X'}^4)^2 + ({X'}^5)^2
\end{array}
\right).
\ee
Then 
\be
g := det(g_{ab}) = \left(({X'}^4)^2 + ({X'}^5)^2\right)(\dot{X})^2 + \left(\dot{X}^4{X'}^5\right)^2.
\ee
We can now reduce the world-sheet action as follows
\be
\int_{\Sigma} \sqrt{-g}\phi = \int_{\gamma} d\tau \phi \sqrt{-\left(({X'}^4)^2 + ({X'}^5)^2\right)(\dot{X})^2 - \left(\dot{X}^4{X'}^5\right)^2}
\ee
Classically this action is equivalent to the action
\be
\frac{1}{2}\int d\tau \left\{\eta^{-1}\left((\dot{X})^2 + \frac{({X'}^5)^2}{({X'}^4)^2 + ({X'}^5)^2}(\dot{X}^4)^2\right) - \eta\phi^2 \(({X'}^4)^2 + ({X'}^5)^2\) \right\}.
\ee
where we have introduced an auxiliary tetrad field $\eta$ on the world-line. We may eliminate $\dot{X}^4$ in favour of the conserved conjugate momentum $P = \frac{\partial L}{\partial \dot{X}^4} = \eta^{-1}\frac{({X'}^5)^2}{({X'}^4)^2 + ({X'}^5)^2}\dot{X}_4$ by making a Legendre transformation with respect to $X^4$. We then get the action
\be
\int d\tau \left\{L(X^{\mu},\dot{X}^{\mu},P) - P \dot{X}^4\right\} = \frac{1}{2}\int d\tau \left\{\eta^{-1}(\dot{X})^2 - \eta\(\frac{({X'}^4)^2 + ({X'}^5)^2}{({X'}^5)^2}P^2 + \phi^2 \(({X'}^4)^2 + ({X'}^5)^2\)\)\right\}
\ee
and this is classically equivalent to the particle action used in \cite{Manton} if we identify $P$ with the rest mass of the dyon in the absence of the scalar field. We now see that we should define the four-dimensional scalar fields as $\Phi^a = R\phi^a$ in order to get the action in \cite{Manton}. The sixth scalar should then be defined as $\Phi^6 \sim R B_{45}$ in order to have the same mass-dimension on all the scalars. (The normalization constant can be determined from the (2,0) supersymmetry, and the requirement that $\Phi^{{\cal{A}}}$ (${\cal{A}} = 1,...,6$) should transform as a 6-vector under the R-symmetry $SO(6)$.) We now see that the term $R^2 B_{45} \sim R \Phi^6$ becomes irrelevant compared to the other scalars, as $R\rightarrow 0$.

Charge quantization is a quantum effect. In the classical limit the electric charge takes a continuum of values. From Dirac quantization of the charges it follows that if we take the limit $\hbar \rightarrow 0$ while we keep both the electric and magnetic charges $nq$ and $mg$ fixed, then we must take $n,m \rightarrow \infty$, and hence the charges will take a continuum of values. If we at the same time take the weak coupling limit $q\rightarrow 0$ as we let $\hbar \rightarrow 0$ then we may $m$ fixed and finite. When taking any of these limits we may thus continuously rotate the strings relative to each other. Two strings of types $(n,m)$ and $(n',m)$ will correspond to two dyons with same magnetic charges, but with different electric charges. 

We thus regain the action that was considered in \cite{Manton}. From this action the Taub-NUT moduli space was obtained there. The corresponding exact metric and moduli space was obtained in \cite{Hitchin}. Our six dimensional theory does not reproduce that exact moduli space in four dimensions. It would be very interesting if one could determine the exact moduli space metric of the six dimensional theory from knowing the asymptotic behaviour as well as the exact moduli space of the four dimensional theory. 

Thus far almost everything we know about these theories have been deduced from string theory. But if we want to learn something new about string theory by studying these six-dimensional theories (which of course is our ultimate goal), then it seems to be more logical to see what can learned about them by using the fact that they should reduce to superYang-Mills theories in four dimensions.

\end{document}